\documentclass[onecolumn,amsmath,amssymb,superscriptaddress,nofootinbib,11pt,a4paper]{revtex4}
\usepackage{booktabs}
\usepackage{caption}
\usepackage{multirow}
\usepackage{array}
\pdfoutput=1
\usepackage[T1]{fontenc}
\usepackage{CJK}
\usepackage{xcolor}
\usepackage{amsfonts}
\usepackage{epstopdf}
\usepackage{wrapfig} 
\usepackage{subcaption}
\usepackage{graphicx}  
\usepackage{dcolumn}   
\usepackage{bm}
\usepackage{float}
\usepackage[T1]{fontenc}
\usepackage{CJK}
\usepackage{xcolor}
\usepackage{amsfonts}
\usepackage{epstopdf}
\usepackage{wrapfig} 
\usepackage{subcaption} 
\usepackage{graphicx}  
\usepackage{dcolumn}   
\usepackage{bm}
\usepackage{float}
\usepackage[utf8]{inputenc}
\usepackage[T1]{fontenc}

\begin{document}

\title{Repetitive Penrose process in Kerr-de Sitter black holes}

\author{Ke Wang}
\email{kkwwang2025@163.com}
\affiliation{School of Material Science and Engineering, Chongqing Jiaotong University, \\Chongqing 400074, China}

\author{Xiao-Xiong Zeng\footnote{Electronic address: xxzengphysics@163.com.  (Corresponding author)}}
\affiliation{College of Physics and Electronic Engineering, Chongqing Normal University, \\Chongqing 401331, China}

\begin{abstract}
{Recently, references \cite{8,4} found that the repetitive Penrose process cannot extract all the extractable rotational energy of a Kerr black hole, and reference \cite{5} found that the repetitive electric Penrose process cannot extract all the electrical energy of a Reissner-Nordström (RN) black hole. In this paper, we intend to study the repetitive Penrose process in the Kerr-de Sitter (Kerr-dS)  black hole. We will explore influences of the cosmological parameter on the repetitive Penrose process. The results show that, apart from the comparable features documented earlier, the Kerr-dS black hole yields a higher energy return on investment (EROI) and single-extraction energy capability compared to the Kerr black hole. Specifically, the larger the cosmological parameter, the stronger the EROI and the single-extraction energy capability. Furthermore, we also find that at a lower decay radius, the Kerr black hole exhibits a higher energy utilization efficiency (EUE) and more extracted energy after the repetitive Penrose process is completed. However, at a higher decay radius, the situation is reversed, i.e., the Kerr-dS black hole exhibits a higher EUE and more extracted energy, which is due to the existence of stopping condition of the iteration.
}
\end{abstract}

\maketitle
 \newpage
\section{Introduction}
Extracting energy from black holes has long been a topic of great interest to physicists. In 1969, Penrose proposed a method for extracting energy from a rotating black hole \cite{9}. A particle with mass $\mu_{0}$ and energy $E_{0}$ falls from infinity and decays within the ergosphere into two particles, $\mu_{1}$ and $\mu_{2}$. The particle $\mu_{1}$ with negative energy $E_{1}$ falls into the event horizon of the black hole, while the particle $\mu_{2}$ escapes to infinity carrying more energy $E_{2}$ than the original particle. Thus, rotational energy $E_{2}-E_0$ is extracted from the black hole. However, Bardeen, Wald, and others pointed out that the original Penrose process lacks a feasible initiation mechanism. In order to extract  energy from the black holes, the velocity of the outgoing particle should exceed half the speed of light \cite{10,11}. Building on Penrose’s pioneering work, scientists have discovered many other energy extraction mechanisms, such as the magnetic Penrose process \cite{12,13}, the collisional Penrose process \cite{14}, superradiant scattering \cite{15}, the Blandford–Znajek mechanism \cite{16}, the magnetohydrodynamic Penrose process \cite{17}, the Bañados–Silk–West mechanism \cite{18}, the possible energy-extraction Penrose-like mechanism which violates the null energy condition\cite{32,33}, and the Comisso–Asenjo mechanism \cite{19}. In particular, based on the Comisso–Asenjo process,  the study of magnetic reconnection mechanisms, as well as the power and efficiency of energy extraction via such mechanisms, have been investigated extensively\cite{6,7,26,27,28,29,30,31}.

On the other hand, Misner, Thorne, and Wheeler proposed a repetitive Penrose process \cite{20}. They envisioned an advanced civilization constructing a rigid framework around a black hole and building a vast city on it. Daily, trucks would transport one million tons of garbage from the city to a dumping ground. At the dump, the garbage would be loaded into shuttles, which would then be dropped one after another toward the center of the black hole. In terms of the Penrose process, this means repeating the decay in the same manner, one after another, into the black hole. However, Ruffini et al. \cite{8} found very recently that such a process would violate energy conservation. This is because the effect of the irreducible mass \cite{21,2,3} was not taken into account in the process. For each iteration, the mass and spin of the black hole were not updated, treating the iterative process as a linear one. In light of this, references \cite{8,4} proposed a modified repetitive Penrose process, wherein for each iteration, the updated mass and spin of the black hole are used to extract energy, while also accounting for the new irreducible mass. This renders the iterative process nonlinear. They discovered that in the repetitive Penrose process, it is impossible to extract all the extractable energy of a Kerr black hole, and the reduction in extractable energy is primarily converted into irreducible mass. Currently, the work of references \cite{8,4} has been extended to the electric Penrose process \cite{5}, where it was found that the repetitive electric Penrose process cannot extract all the electrical energy of a RN black hole.

Inspired by the works in references \cite{8,4}, this paper will investigate the repetitive Penrose process in Kerr-dS spacetime. Kerr-dS spacetime is an exact solution to the Einstein field equations with a positive cosmological constant. It describes a rotating black hole embedded in a de Sitter universe. Currently, a growing body of research focuses on a positive cosmological constant, such as  cosmic inflation \cite{22}, phase transitions \cite{23}, thermodynamics \cite{24},  dS/CFT correspondence \cite{25}, and so forth. In Kerr-dS spacetime, in addition to the impossibility of reaching a state of zero extractable energy through the repetitive Penrose process, we will also explore the influence of the cosmological parameter on the repetitive Penrose process. In particular, we will compare our results with those in  the Kerr black hole in terms of the  EROI, single-extraction energy capability, EUE, and the variation in total extracted energy. The results indicate that, in addition to exhibiting previously observed similar phenomena, Kerr-dS spacetime possesses a higher EROI and single-extraction energy capability compared to Kerr spacetime. The larger the cosmological parameter, the stronger the EROI and single-extraction energy capability. In the case of a lower decay radius, after the completion of the repetitive Penrose process, the Kerr black hole exhibits a higher EUE and yields more extracted energy. However, in the case of a higher decay radius, the situation is reversed. That is, the Kerr-dS black hole exhibits a higher EUE and yields more extracted energy.

The remainder of this paper is organized as follows. In Section 2, we will introduce the Penrose process in Kerr-dS black holes. In Section 3, we will investigate the repetitive Penrose process in Kerr-dS black holes, with particular focus on exploring the influence of the cosmological parameter on this process. We present our summary in Section 4. Throughout the paper, we will employ geometric units $(c=G=1)$.

\section{The Penrose Process in Kerr-dS Black Holes}
In Boyer-Lindquist coordinates and geometric units, the line element for the Kerr-dS metric is given by \cite{1}
\begin{equation}
ds^{2} = g_{tt} dt^{2} + g_{rr} dr^{2} + g_{\theta\theta} d\theta^{2} + 2g_{t\phi} dt d\phi +g_{\phi\phi} d\phi^{2},
\end{equation}
where the metric components are given by
\begin{equation}
\begin{aligned}
g_{tt} = -\frac{\Delta_{r} - \Delta_{\theta} a^{2} \sin^{2} \theta}{\rho^{2} \Sigma^{2}},  g_{t\phi} = -\frac{\Delta_{\theta} (r^{2} + a^{2}) - \Delta_{r}}{\rho^{2} \Sigma^{2}} a \sin^{2} \theta,\\
g_{rr} = \frac{\rho^{2}}{\Delta_{r}},g_{\theta\theta} = \frac{\rho^{2}}{\Delta_{\theta}}, g_{\phi\phi} = \frac{\Delta_{\theta} (r^{2} + a^{2})^{2} - \Delta_{r} a^{2} \sin^{2} \theta}{\rho^{2} \Sigma^{2}} \sin^{2} \theta.
\end{aligned}
\end{equation}
The metric functions are
\begin{equation}
\rho^{2} = r^{2} + a^{2} \cos^{2} \theta,  \Sigma = 1 + \frac{1}{3} \Lambda a^{2},  \Delta_{r} = (r^{2} + a^{2}) \left(1 - \frac{1}{3} \Lambda r^{2}\right) - 2Mr,  \Delta_{\theta} = 1 + \frac{1}{3} \Lambda a^{2} \cos^{2} \theta.
\end{equation}
Here, $a$ is the spin of the black hole, $M$ is the mass of the black hole, and $\Lambda$ is the positive cosmological constant. When $\Lambda=0$, the metric reduces to the Kerr metric. The horizons of the black hole are determined by $\Delta_{r}=0$. This equation has four roots in total, ordered from smallest to largest as $r_{--}$, $r_{-}$, $r_{+}$, and $r_{c}$. Among these, $r_{--}$ is always less than zero and is discarded, $r_{-}$ is the Cauchy horizon, $r_{+}$ is the event horizon, and $r_{c}$ is the cosmological horizon. The corresponding solutions satisfy $r_{--} < r_{-} \leq r_{+} \leq r_{c}$. The boundaries of the ergosphere are located at $g_{tt} = 0$. In general, this also yields four roots, ordered from smallest to largest as $r_{E--}$, $r_{E-}$, $r_{E}$, and $r_{Ec}$. Here, $r_{E--}$ is always less than zero and is discarded, $r_{E-}$ is the inner ergosphere boundary, $r_{E}$ is the event horizon ergosphere boundary, and $r_{Ec}$ is the cosmological horizon ergosphere boundary. In this paper, we will extract energy  within the ergosphere of the event horizon. Under general conditions preserving the existence of the ergosphere, the hierarchy of the aforementioned horizons and ergosphere boundaries satisfies
\begin{equation}
r_{E-}\leq r_{-}\leq r_{+} \leq r_{E} \leq r_{Ec}\leq r_{c}.
\end{equation}
This inequality constrains the choices of the cosmological parameter and the spin. In this paper, to maximize energy extraction, we will consider the repetitive Penrose process for extremal Kerr-dS black holes on the equatorial plane. That is, for a given cosmological parameter, the initial spin corresponds to the extremal black hole case, i.e., satisfying $r_+=r_-$\footnote{What needs to be emphasized here is that by "extremal black hole," we refer to the initial black hole being extremal, and not that it remains extremal after iterations of the repetitive Penrose process. This is consistent with the approach in reference \cite{4}, where they consider an initial extremal Kerr black hole with $a=M$, which does not remain extremal after subsequent iterations.}. Here, $\Lambda M^2$ can range from 0 to 0.099555, and $a/M$ ranges from 1 to 1.04075.

\begin{figure}[!h]
  \centering
  \setlength{\tabcolsep}{2pt}
  \begin{tabular}{ccc}
    \includegraphics[width=0.32\linewidth]{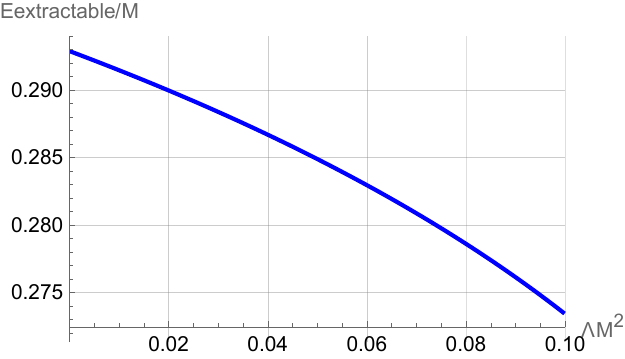} &
    \includegraphics[width=0.32\linewidth]{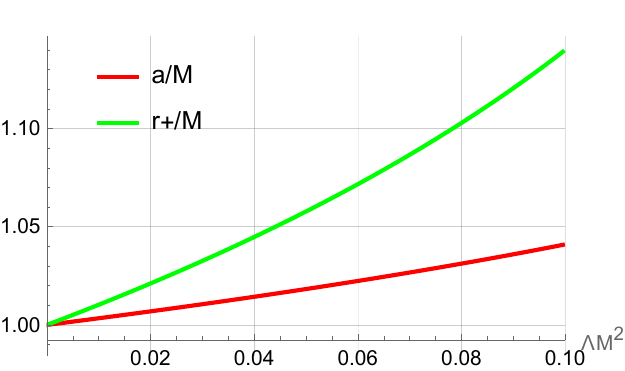} &
    \includegraphics[width=0.32\linewidth]{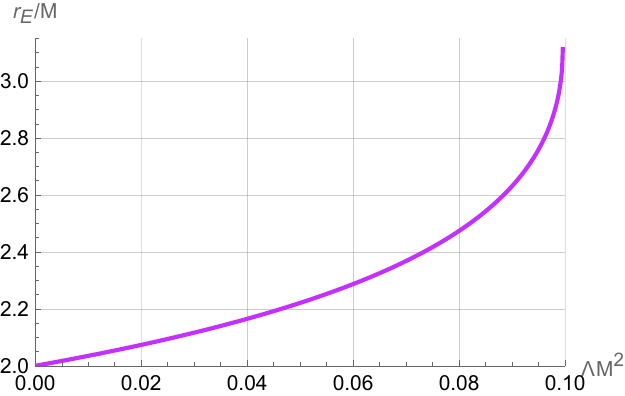}
  \end{tabular}
  \caption{Variation of $E_{extractable}/M$, $a/M$, $r_{+}/M$, and $r_{E}/M$ for an extremal black hole with respect to $\Lambda M^2$.}
\label{fig:1}
\end{figure}

Within the event horizon, in addition to the mass $M$ itself, there exists a local positive energy contribution due to the cosmological constant \cite{34}. A local, continuous, positive-definite effective mass function can be expressed as \cite{34}
\begin{equation}
M_{loc}(r)= M + \frac{\Lambda r}{6}\left(r^2 +a^2\right).
\end{equation}
Thus, the irreducible mass can be written as \cite{34}
\begin{equation}
M_{irr} =\sqrt{ \frac{1}{2} M_{loc,H}(M_{loc,H} + \sqrt{M_{loc,H}^{2} - a^{2}})}.
\end{equation}
Here, $M_{loc,H}$ denotes the value of $M_{loc}(r)$ at the event horizon.
Then the extractable energy can be expressed as
\begin{equation}
E_{extractable}= M_{loc,H}-M_{irr},
\end{equation}
which represents the maximum energy theoretically extractable from a Kerr-dS black hole. In Fig. \ref{fig:1}, we plot the maximum extractable energy, the spin of the extremal black hole, and the event horizon along with its ergosphere boundary for an extremal black hole as functions of $\Lambda M^2$. It can be observed that as $\Lambda M^2$ grows, the spin of the extremal black hole, its event horizon, and its ergosphere boundary all increase monotonically, while the maximum extractable energy gradually decreases.

The fundamental equations of the Penrose process are given by four-momentum conservation. The conservation equations for energy, angular momentum, and radial momentum can be expressed as
\begin{equation}
\begin{cases}
\hat{E}_{0}=\tilde{\mu}_{1}\hat{E}_{1}+\tilde{\mu}_{2}\hat{E}_{2}, \\
\hat{p}_{\phi 0}=\tilde{\mu}_{1}\hat{p}_{\phi 1}+\tilde{\mu}_{2}\hat{p}_{\phi 2},\\
\hat{p}_{r0}=\tilde{\mu}_{1}\hat{p}_{r1}+\tilde{\mu}_{2}\hat{p}_{r2},
\end{cases}\label{8}
\end{equation}
where
\begin{equation}
\hat{E}_{i}={E}_{i}/{\mu}_{i},\hat{p}_{\phi i}=p_{\phi i}/(\mu_{i}M),\hat{p}_{ri}=p_{ri}/\mu_{i},\tilde{\mu}_{i}=\mu_{i}/\mu_{0}.
\end{equation}
Here, $\mu_{i}$ is the mass of particle $i$.
According to the normalization condition, the four-momentum of a particle satisfies
\begin{equation}
g^{\mu\nu} p_\mu p_\nu = -\mu^2,
\end{equation}
where $p_t = -E$, and $g^{\mu\nu}$ is the inverse metric. For motion on the equatorial plane, $\theta = \pi/2$, so $p_\theta = 0$. When a particle is at a turning point, its radial momentum $p_r = 0$, and thus the normalization condition simplifies to
\begin{equation}
g^{tt} E^2 - 2 g^{t\phi} E {p}_{\phi} + g^{\phi\phi} {p}_{\phi}^2 = -\mu^2.\label{11}
\end{equation}
The effective potential for the radial motion of a particle is defined as the value of $E$ when $p_r = 0$. Therefore, solving equation \eqref{11} yields the effective potentials for particles 0, 1, and 2
\begin{equation}
\hat{V}^\pm_i = \frac{ g^{t\phi} \hat{p}_{\phi i} M \mp \sqrt{ (g^{t\phi})^{2} \hat{p}_{\phi i}^{2} M^{2} - g^{tt} \left( g^{\phi\phi} \hat{p}_{\phi i}^{2} M^{2} + 1 \right) } }{ g^{tt} }.
\end{equation}
If $p_r \neq 0$, then according to the normalization condition, we obtain
\begin{equation}
\hat{p}_{ri}^2 = - \frac{g^{tt}}{g^{rr}} (\hat{E}_i - \hat{V}_i^+) (\hat{E}_i - \hat{V}_i^-).\label{pri}
\end{equation}
References \cite{4,5} demonstrate that to satisfy the optimal conditions for maximum energy extraction, the three particles must not violate the timelike geodesic condition, and their radial momenta must be zero at the decay position.

\begin{figure}[!h]
  \centering
  \setlength{\tabcolsep}{2pt}
  \begin{tabular}{c}
    \includegraphics[width=0.5\linewidth]{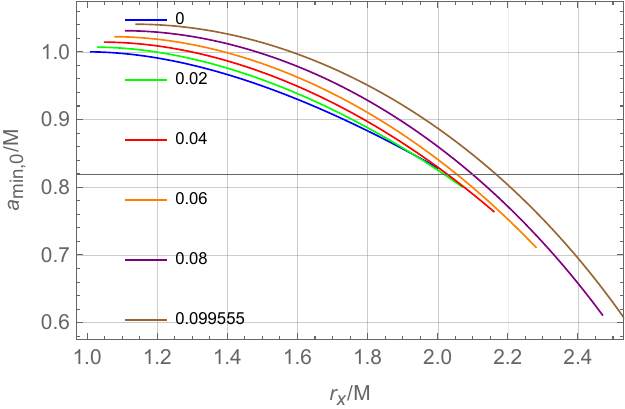}
  \end{tabular}
  \caption{Variation of $\hat{a}_{\min,0}$ with respect to $\hat r_x$ for different values of $\Lambda M^2$.}
  \label{fig:2}
\end{figure}

\begin{figure}[!h]
  \centering
  \setlength{\tabcolsep}{2pt}
  \begin{tabular}{c}
    \includegraphics[width=0.5\linewidth]{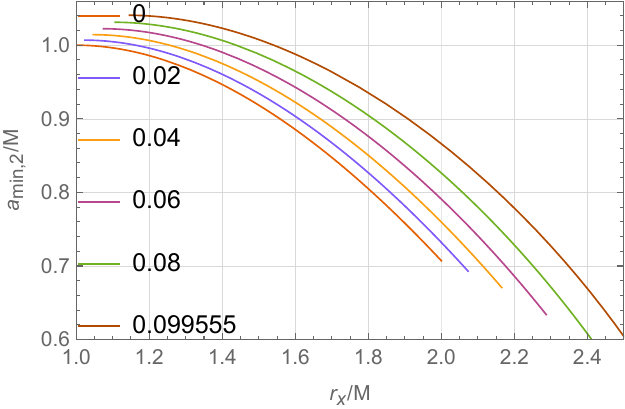}
  \end{tabular}
  \caption{Variation of $\hat{a}_{\min,2}$ with respect to $\hat r_x$ for different values of $\Lambda M^2$.}
  \label{fig:3}
\end{figure}

\begin{figure}[!h]
  \centering
  \setlength{\tabcolsep}{2pt}
  \begin{tabular}{c}
    \includegraphics[width=0.5\linewidth]{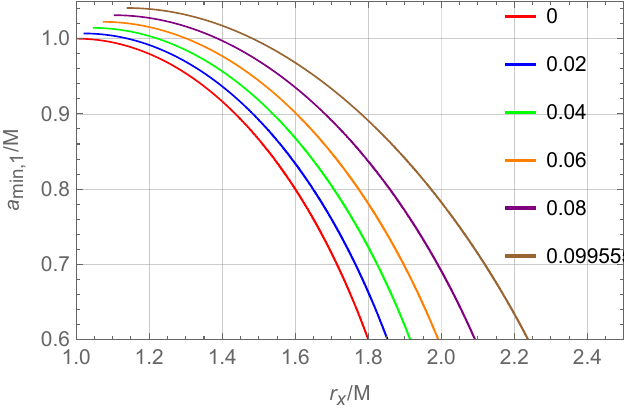}
  \end{tabular}
  \caption{Variation of $\hat{a}_{\min,1}$ with respect to $\hat r_x$ for different values of $\Lambda M^2$.}
  \label{fig:4}
\end{figure}

\begin{figure}[!h]
  \centering
  \setlength{\tabcolsep}{2pt}
  \begin{tabular}{ccc}
    \includegraphics[width=0.32\linewidth]{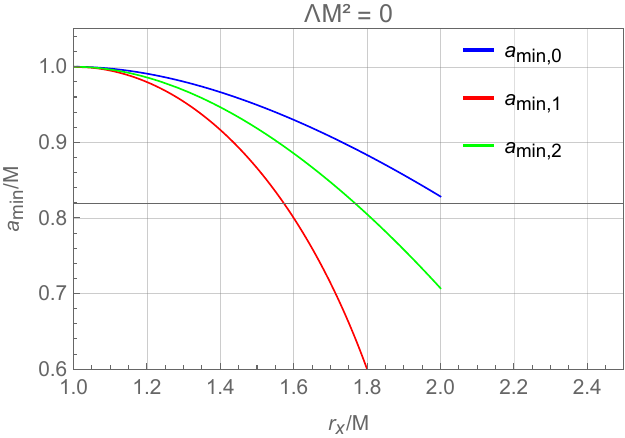} &
    \includegraphics[width=0.32\linewidth]{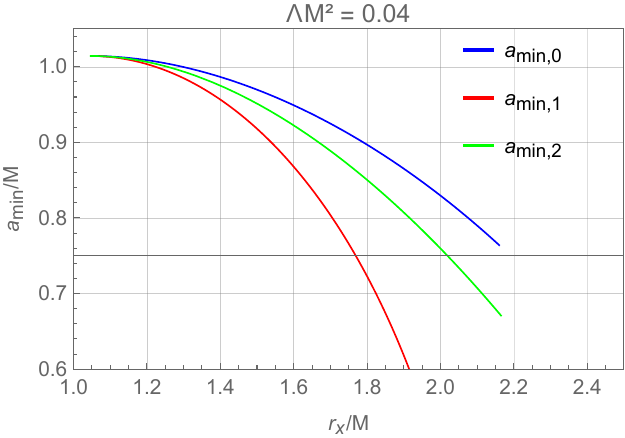} &
    \includegraphics[width=0.32\linewidth]{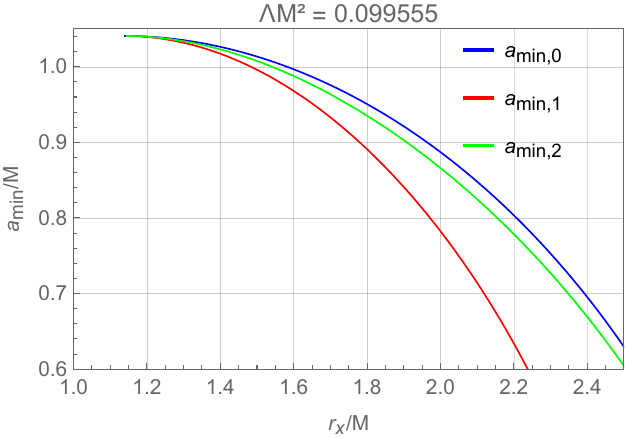} 
  \end{tabular}
  \caption{Variation of $\hat{a}_{\min}$ for the three particles with respect to $\hat r_x$ for different values of $\Lambda M^2$.}
  \label{fig:5}
\end{figure}

Indeed, this relation holds for general metrics as well. Since $g^{tt} < 0$ and $g^{rr} > 0$, based on $\hat{p}_{ri}^2 > 0$, we have
$\hat{E}_i \geq \hat{V}_i^+$ or $\hat{E}_i \leq \hat{V}_i^-$. Secondly, for the timelike geodesic condition, it is required that $dt/d\tau > 0$. The expression for $dt/d\tau$ can be obtained from the definition of four-momentum, $p_\mu = \mu \frac{dx_\mu}{d\tau}$, that is
\begin{equation}
\frac{dt}{d\tau} = -g^{tt} \hat{E} + g^{t\phi} M \hat{p}_\phi.\label{timelike}
\end{equation}
Evaluating $dt/d\tau$ at the effective potential yields
\begin{equation}
\left. \frac{dt}{d\tau} \right|_{\hat{E}_i = \hat{V}_i^\pm} = \pm \sqrt{ (g^{t\phi})^{2} \hat{p}_{\phi i}^{2} M^{2} - g^{tt} \left( g^{\phi\phi} \hat{p}_{\phi i}^{2} M^{2} + 1 \right) }.
\end{equation}
Here, the expression under the square root is greater than or equal to zero. However, since $dt/d\tau \neq 0$, the expression under the square root is in fact strictly greater than zero. When $\frac{dt}{d\tau}|_{\hat{E}_i = \hat{V}_i^-}$, we have $\frac{dt}{d\tau} < 0$, which leads to a contradiction. Therefore, only the branch $\hat{E}_i \geq \hat{V}_i^+$ can be taken. For energy extraction, it is required that $\hat{E}_1 < 0$. To satisfy the optimal conditions for maximum energy extraction, $\hat{E}_1$ should be as negative as possible. Combined with the condition $\hat{E}_1 \geq \hat{V}_1^+$, the absolute value of $\hat{E}_1$ can only be maximized when particle 1 is at a turning point, i.e., $\hat{E}_1 = \hat{V}_1^+$. At this point, according to the third equation in formula \eqref{8}, we obtain $p_{r0} = p_{r2}$. Assuming $p_{r0} = p_{r2} \neq 0$, then based on formula \eqref{pri}, we find
\begin{equation}
\begin{cases}
E_{0}-V_{0}^{-}=E_{2}-V_{2}^{-},\\ E_{0}-V_{0}^{+}=E_{2}-V_{2}^{+}.
\end{cases}
\end{equation}
From the conservation of energy and angular momentum, i.e., according to the first two equations in formula \eqref{8}, we obtain 
\begin{equation}
\hat{E}_{1}=\dfrac{g^{t\phi}M}{g^{tt}}\hat{p}_{\phi1}.
\end{equation}
Clearly, this violates the timelike geodesic condition \eqref{timelike}. Therefore, for any metric, in order to satisfy the optimal conditions for maximum energy extraction, the radial momenta of the three particles must be zero at the decay location. At this point, assuming $\hat{E}_{0}$, $\hat{p}_{\phi 1}$, and $\nu = \mu_2/\mu_1$ are known quantities, the fundamental equations of the Penrose process \eqref{8} admit an analytical solution
\begin{equation}
\begin{aligned}
\hat{p}_{\phi 0} &=\frac{ g^{t\phi} \hat{E}_0 +\sqrt{ (g^{t\phi})^2 \hat{E}_0^2 - g^{\phi\phi} (1 + g^{tt} \hat{E}_0^2) } }{ M g^{\phi\phi} }, \\
\hat{E}_1 &= \frac{ g^{t\phi} \hat{p}_{\phi 1} M - \sqrt{ (g^{t\phi})^{2} \hat{p}_{\phi 1}^{2} M^{2} - g^{tt} \left( g^{\phi\phi} \hat{p}_{\phi 1}^{2} M^{2} + 1 \right) } }{ g^{tt} },\\
\tilde{\mu}_1&= \frac{\hat{E}_{0}\hat{E}_{1}g^{tt} - \hat{E}_{1}g^{t\phi}M\hat{p}_{\phi 0} - \hat{E}_{0}g^{t\phi}M\hat{p}_{\phi 1} + g^{\phi\phi}M^{2}\hat{p}_{\phi 0}\hat{p}_{\phi 1} + \sqrt{D}}{\hat{E}_{1}^{2}g^{tt} - 2\hat{E}_{1}g^{t\phi}M\hat{p}_{\phi 1} + g^{\phi\phi}M^{2}\hat{p}_{\phi 1}^{2} + \nu^{2}},\\
\hat{E}_{2} &=\frac{\hat{E}_{0}}{\tilde{\mu}_{2}}-\frac{\hat{E}_{1}}{\nu}, \hat{p}_{\phi 2}=\frac{\hat{p}_{\phi 0}}{\tilde{\mu}_{2}}-\frac{\hat{p}_{\phi 1}}{\nu},\label{13}
\end{aligned}
\end{equation}
where
\begin{equation}
\begin{aligned}
D = & - g^{tt} g^{\phi\phi} M^2 \hat{E}_1^2 \hat{p}_{\phi 0}^2  + (g^{t\phi})^2 M^2 \hat{E}_1^2 \hat{p}_{\phi 0}^2  - g^{tt} g^{\phi\phi} M^2 \hat{E}_0^2 \hat{p}_{\phi 1}^2  + (g^{t\phi})^2 M^2 \hat{E}_0^2 \hat{p}_{\phi 1}^2 \\
& + 2 g^{tt} g^{\phi\phi} M^2 \hat{E}_0 \hat{E}_1 \hat{p}_{\phi 0} \hat{p}_{\phi 1}  - 2 (g^{t\phi})^2 M^2 \hat{E}_0 \hat{E}_1 \hat{p}_{\phi 0} \hat{p}_{\phi 1}  - g^{tt} \hat{E}_0^2 \nu^2  + 2 g^{t\phi} M \hat{E}_0 \hat{p}_{\phi 0} \nu^2 \\
& - g^{\phi\phi} M^2 \hat{p}_{\phi 0}^2 \nu^2.
\end{aligned}
\end{equation}
After each energy extraction, the remaining mass and angular momentum of the black hole are
\begin{equation}
M_{n}=M_{n-1}+\hat{E}_{1,n-1}\mu_{1,n-1},L_{n}=L_{n-1}+\hat{p}_{\phi 1}\mu_{1,n-1}M_{n-1}.
\end{equation}
This will induce certain changes in $\hat{a}=a/M$ and $\hat{\Lambda}=\Lambda M^2$, which are 
\begin{equation}
\Delta\hat{a}_{n-1}=\frac{L_{n}}{M_{n}^{2}}-\frac{L_{n-1}}{M_{n-1}^{2}},\Delta\hat{\Lambda}_{n-1}=\Lambda M_{n}^{2}-\Lambda M_{n-1}^{2}.
\end{equation}
Similarly, according to the horizon equation, the irreducible mass formula, and the expression for extractable energy, $r_+,M_{irr}$ and $E_{ extractable}$ will also change. 
During this process, the extracted energy is
\begin{equation}
E_{ extracted,n}=M_0-M_n.
\end{equation}
The EROI, defined as the ratio of the extracted energy to the total energy of all incident particles from infinity, is given by \cite{5}
\begin{equation}
\xi_n=E_{ extracted,n}/(nE_0).
\end{equation}
The EUE, defined as the ratio of the extracted energy to the difference between the initial and final extractable energy, is given by \cite{5}
\begin{equation}
\Xi_{n}=E_{ extracted,n}/(E_{ extractable,0}-E_{ extractable,n}) .
\end{equation}
In the repetitive energy extraction process, the above iteration cannot continue indefinitely but must satisfy certain conditions. Firstly, the mass deficit must satisfy
\begin{equation}
\mu_{0}-\mu_{1}-\mu_{2}>0.\label{21}
\end{equation}
Secondly, during the iteration process, the condition $\hat{E}_{1}<0$ must be satisfied. Finally, the turning points for particles 0 and 2 must lie on the right side of the peak of their effective potentials, while the turning point for particle 1 must lie on the left side of the peak of its effective potential. The corresponding limiting case is when the classical turning point of each particle coincides precisely with the peak of its respective effective potential, i.e.,
\begin{equation}
\hat{V}_{i}^{+}(\hat{r}_x)=\hat{E}_{i}, {d}\hat{V}_{i}^{+}/{ d}\hat{r}|_{\hat{r}=\hat{r}_{x}}=0,
\end{equation}
where $\hat{r}_x={r}_x/M$ is the dimensionless decay radius. If $\hat{E}_0=1$, the minimum lower bound for the spin of particle 0 lies at the corotating marginally bound orbit of that particle. For a Kerr-dS black hole, the angular velocity and specific energy for a particle in a corotating Keplerian orbit are given by \cite{6}
\begin{equation}
\Omega_K = \frac{-\partial_r g_{t\phi} + \sqrt{(\partial_r g_{t\phi})^2 - (\partial_r g_{tt})(\partial_r g_{\phi\phi})}}{\partial_r g_{\phi\phi}},\hat{\mathcal{E}} = -\frac{g_{tt} + g_{t\phi} \Omega_K}{\sqrt{-g_{tt} - 2g_{t\phi} \Omega_K - g_{\phi\phi} \Omega_K^2}}.\label{23}
\end{equation}
For a marginally bound orbit, $\hat{\mathcal{E}}=1$. Then, the lower spin limit for particle 0 can be obtained by solving equation \eqref{23}. If $\Lambda M^2=0$, the solution to this equation is $\hat{a}_{\min,0}=2\sqrt{\hat{r}_x}-\hat{r}_x$. In Fig. \ref{fig:2}, we plot the variation of $\hat{a}_{\min,0}$ with respect to $\hat r_x$ for different values of $\hat{\Lambda}$. The ranges of the horizontal coordinate differ for different $\hat{\Lambda}$ values due to their distinct event horizon ergospheres. Furthermore, we omit the regions with larger $\hat r_x$ and very small $\hat{a}_{\min}$, as this does not affect the overall trend.
It can be observed that as $\Lambda M^2$ increases, the minimum spin limit for particle 0 also increases, while as $\hat r_x$ increases, the minimum spin limit for particle 0 decreases. If $\hat{E}_0>1$, then the minimum spin limit for particle 2 lies at the corotating photon sphere radius of that particle. The corotating photon sphere radius for a Kerr-dS black hole is given by \cite{7}
\begin{equation}
\hat{r}_{p} = \frac{6\left(3-\hat{a}^{2}\hat{\Lambda}-\sqrt{\hat{a}^{4}\hat{\Lambda}^{2}-42\hat{a}^{2}\hat{\Lambda}+9 }\sin\left(\frac{1}{3}\arcsin\chi\right)\right)}{\left(\hat{a}^{2}\hat{\Lambda}+3\right)^{2}},
\end{equation}
where
\begin{equation}
\chi = \frac{2\hat{a}^{10}\hat{\Lambda}^{4} + 24\hat{a}^{8}\hat{\Lambda}^{3} + 3\hat{a}^{6}\hat{\Lambda}^{3} + 108\hat{a}^{6}\hat{\Lambda}^{2} + 297\hat{a}^{4}\hat{\Lambda}^{2} + 216\hat{a}^{4}\hat{\Lambda} - 891\hat{a}^{2}\hat{\Lambda} + 162\hat{a}^{2} - 81}{3\left(\hat{a}^{4}\hat{\Lambda}^{2} - 42\hat{a}^{2}\hat{\Lambda} + 9\right)^{3/2}}.
\end{equation}
The minimum spin limit for particle 2 can be obtained by inversely solving the above formula. In Fig. \ref{fig:3}, we plot the variation of $\hat{a}_{\min,2}$ with respect to $\hat r_x$ for different values of $\hat{\Lambda}$.
It can be observed that the variation pattern of the minimum spin limit for particle 2 is very similar to that for particle 0. However, under the same parameters, $\hat{a}_{\min,2} < \hat{a}_{\min,0}$, which may not be immediately obvious, and we will compare them in detail later. For the turning point of particle 1 to exist, the discriminant under the square root in the second formula of equation \eqref{13} must be positive. In fact, this corresponds to $\hat r_x$ being greater than the radius of the black hole's event horizon. The critical case is $\hat r_x = \hat r_+$, which defines the minimum spin limit for particle 1. Similarly, in Fig. \ref{fig:4}, we plot the variation of $\hat{a}_{\min,1}$ with respect to $\hat r_x$ for different values of $\hat{\Lambda}$.
It can be observed that the variation pattern of the minimum spin limit for particle 1 is quite similar to that for particles 0 and 2. However, under identical parameters, $\hat{a}_{\min,1}$ is the lowest. To more clearly distinguish the magnitudes among $\hat{a}_{\min,0}$, $\hat{a}_{\min,1}$, and $\hat{a}_{\min,2}$, we plot in Fig. \ref{fig:5} the variation of $\hat{a}_{\min}$ for the three particles with respect to $\hat r_x$ for different values of $\Lambda M^2$.
Figure \ref{fig:5} clearly illustrates the fact that $\hat{a}_{\min,1} < \hat{a}_{\min,2} < \hat{a}_{\min,0}$. Therefore, the lower spin limit for stopping the iteration is determined by particle 0. It is worth noting here that this lower spin limit for stopping the iteration is not fixed, because with each iteration, the mass $M$ decreases, and $\Lambda M^2$ correspondingly decreases. Consequently, as can be inferred from Fig. \ref{fig:2}, the lower spin limit for stopping the iteration slightly decreases with each iteration, but it will not fall below that of the Kerr black hole.

\section{The Repetitive Penrose Process in Kerr-dS Black Holes}
\begin{figure}[!h]
  \centering
  \setlength{\tabcolsep}{2pt}
  \begin{tabular}{cc}
    \includegraphics[width=0.45\linewidth]{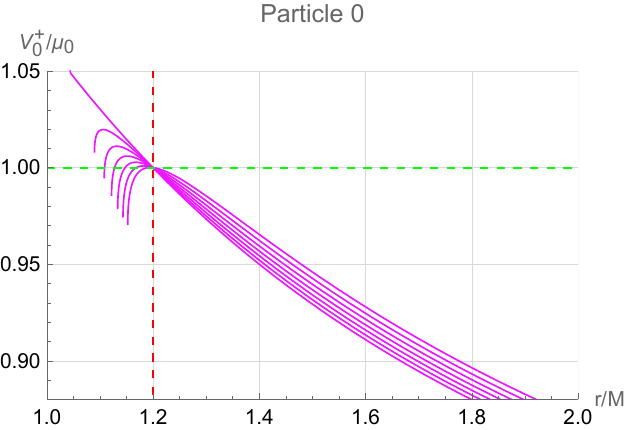} &
    \includegraphics[width=0.45\linewidth]{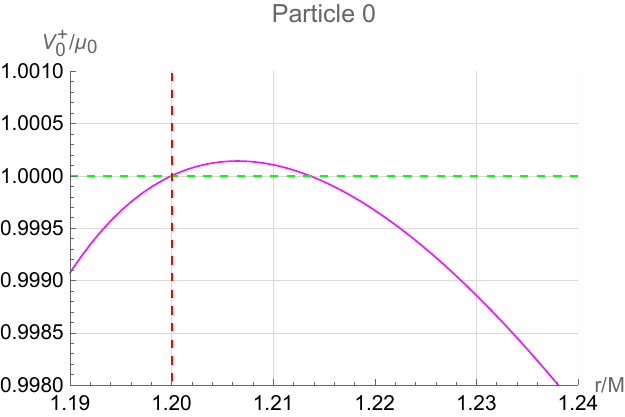} 
  \end{tabular}
  \caption{Left: Magenta solid lines represent the curves of the effective potential versus $\hat{r}$ for iteration numbers $n$ from 0 to 6. The green horizontal dashed line indicates the energy $\hat{E}_0$ of particle 0, and the red vertical dashed line marks the decay radius. Right: The curve of the effective potential versus $\hat{r}$ after forcibly performing the 7th iteration.}
  \label{fig:6}
\end{figure}

In this section, we focus on studying the repetitive Penrose process. The results in reference \cite{4} show that for the case  $\hat{E}_0 > 1$, the EROI is lower than the case $\hat{E}_0 = 1$. Therefore, we choose $\hat{E}_0 = 1$ to maximize the EROI. Following reference \cite{4}, we take $\hat{p}_{\phi 1} = -19.434$, $\nu = \mu_2/\mu_1 = 0.78345$,  $\hat{r}_x = 1.2$, and we set $\mu_0 = 10^{-2}M$. We present the results in Table \ref{tab:1}, with the initial cosmological parameter taken as $\hat{\Lambda} = 0.04$.
\begin{table}[htbp]
\centering
\small 
\caption{Repetitive Penrose process with initial $\hat{\Lambda}=0.04$.} 
\label{tab:1}
\begin{tabular}{ccccccccccc}
\hline
$n$ & $\frac{M_n}{M_0}$ & $\hat{a}_n$ & $\frac{\mu_{1,n}}{\mu_0}$ & $\frac{E_{1,n}}{\mu_0}$ & $\frac{E_{extractable,n}}{M_0}$ &  $\frac{E_{extracted,n}}{M_0}$ & $\frac{M_{{irr,n}}}{M_0}$ & $\xi_n$ & $\Xi_n$ &$\hat{a}_{min,0,n}$ \\
\hline
0 & 1.000000 & 1.014325 & 0.020331 & -0.147712 & 0.286699 & 0.000000 & 0.728069 & 0.000000 & 0.000000 & 1.008737 \\
1 & 0.998523 & 1.013366 & 0.020248 & -0.147710 & 0.271603 & 0.001477 & 0.742935 & 0.147712 & 0.097850 & 1.008681 \\
2 & 0.997046 & 1.012418 & 0.020163 & -0.147708 & 0.265308 & 0.002954 & 0.748231 & 0.147711 & 0.138103 &
1.008625 \\
3 & 0.995569 & 1.011483 & 0.020076 & -0.147707 & 0.260487 & 0.004431 & 0.751921 & 0.147710 & 0.169058 & 
1.008568 \\
4 & 0.994092 & 1.010560 & 0.019986 & -0.147705 & 0.256442 & 0.005908 & 0.754765 & 0.147709 & 0.195276 &
1.008513 \\
5 & 0.992615 & 1.009651 & 0.019893 & -0.147704 & 0.252901 & 0.007385 & 0.757060 & 0.147709 & 0.218518 &
1.008457 \\
6 & 0.991138 & 1.008756 & 0.019797 & -0.147703 & 0.249724 & 0.008862 & 0.758957 & 0.147708 & 0.239688 &
1.008401 \\
\hline
\end{tabular}
\end{table}
All data in Table \ref{tab:1} satisfy the iteration conditions. For example, according to the mass deficit formula \eqref{21}, we obtain $\tilde{\mu}_{1}<1/(1+\mu_{2}/\mu_{1})=0.56$. Additionally, the condition $\hat{E}_{1}<0$ is also satisfied. The last column represents the lower spin limit for stopping the iteration after each iteration, showing that this limit slightly decreases with each iteration. At $n=6$, the iteration has stopped. If we forcibly continue the iteration, we find $\hat{a}_7=1.007876$ and $\hat{a}_{\min,0,7}=1.008345$, which no longer satisfy the iteration conditions. For clarity, we plot the variation of the effective potential for particle 0 during each iteration in Fig. \ref{fig:6}.
From Fig. \ref{fig:6}, it can be observed that as the iteration number increases, the peak of the effective potential decreases. The left plot shows that the turning point of particle 0 lies to the right of the peak of the effective potential. The right plot illustrates that if we forcibly impose a solution below the lower spin limit for stopping the iteration, the turning point of particle 0 would lie to the left of the peak of the effective potential. From Table \ref{tab:1}, it can be seen that it is impossible to extract all the extractable energy from a Kerr-dS black hole. The reduction in extractable energy is partly converted into extracted energy, while the majority flows into irreducible mass. Based on the EUE, it is shown that 23.97$\%$ of the reduction in extractable energy is converted into extracted energy. Unlike in the previous discussion, the reduction in extractable energy is not equal to the sum of the extracted energy and the increased irreducible mass. Moreover, the sum of the initial extractable energy and the initial irreducible mass exceeds the black hole mass. This discrepancy arises due to the influence of local positive energy contributions from spin and the cosmological constant. The definitions of extractable energy and irreducible mass we employ here fully account for these local positive energy effects. The results indicate that reducing the black hole's spin cannot extract all the corresponding rotational energy, which is limited by the nonlinear increase in irreducible mass. Furthermore, after the repetitive Penrose process terminates, there remains $0.249724M$ of extractable energy, suggesting that a considerable amount of energy still awaits extraction through other means.

\begin{figure}[!h]
  \centering
  \setlength{\tabcolsep}{2pt}
  \begin{tabular}{ccc}
    \includegraphics[width=0.32\linewidth]{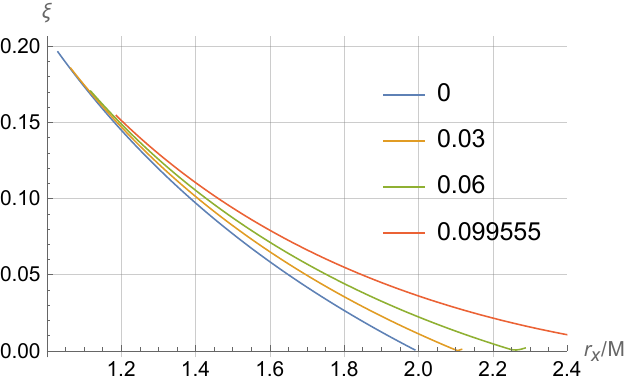} &
    \includegraphics[width=0.32\linewidth]{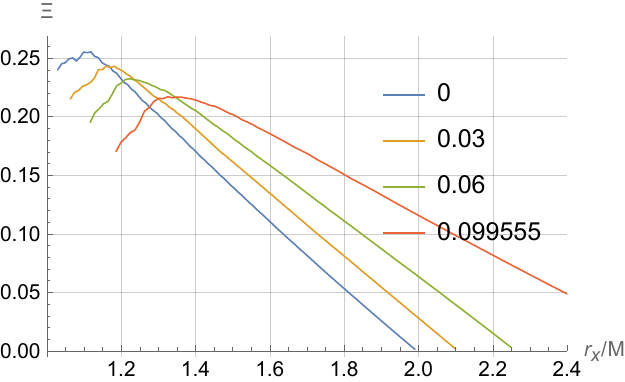} &
    \includegraphics[width=0.32\linewidth]{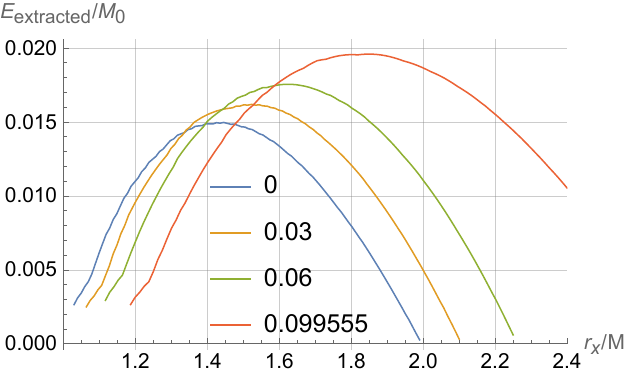} 
  \end{tabular}
  \caption{Variation of $\xi$, $\Xi$, and $E_{extracted}/M_0$ with respect to $\hat{r}_x$ after the iterative termination of the repetitive Penrose process, for different initial values of $\hat{\Lambda}$.}
  \label{fig:7}
\end{figure}

In Fig. \ref{fig:7}, we plot the variation of the EROI $\xi$, the EUE $\Xi$, and the extracted energy $E_{extracted}/M_0$ with respect to the decay radius $\hat{r}_x$ after the iterative termination of the repetitive Penrose process, for different initial values of $\hat{\Lambda}$. All other parameters are the same as those in Table \ref{tab:1}, except for the initial $\hat{\Lambda}$ and the varying decay radius.
It should be emphasized that because the iteration sequence of this process is discrete, the curves exhibit oscillatory behavior, and we have applied smoothing to the curves. Therefore, specific numerical values have slight errors, but the overall trends remain consistent. From the left panel of Fig. \ref{fig:7}, it is clearly observed that, at the same decay radius, Kerr-dS spacetime yields a higher EROI than the Kerr spacetime, and the larger the cosmological parameter, the greater the value of $\xi$. For the same cosmological parameter, a larger decay radius results in a smaller value of $\xi$. Although we present the value of $\xi$ after iterative termination, as shown by $\xi_n$ in Table \ref{tab:1}, the value of $\xi$ for each energy extraction is nearly identical, which will also be evident in the subsequent Table \ref{tab:2}. In other words, at the same decay radius, Kerr-dS spacetime possesses a higher single-extraction energy capability than Kerr spacetime, and a larger cosmological parameter strengthens this single-extraction capability. For the same cosmological parameter, a larger decay radius weakens the single-extraction capability. From the middle panel of Fig. \ref{fig:7}, it can be seen that at a smaller decay radius, for the same decay radius, a larger cosmological parameter leads to a lower EUE. This indicates that a larger portion of the reduction in extractable energy is converted into irreducible mass, leaving less to become extracted energy, which is reflected in the right panel of Fig. \ref{fig:7}. However, at a larger decay radius, the opposite situation occurs. To facilitate understanding of this phenomenon, we present the repetitive Penrose process for four initial cosmological parameter  cases at a lower decay radius ($\hat{r}_x=1.2$) in Table \ref{tab:2}. All other parameters are the same as those in Table \ref{tab:1}.
\begin{table}[htbp]
\centering
\caption{Repetitive Penrose process for four initial cosmological constant cases with $\hat{r}_x=1.2$.}
\label{tab:2}
\begin{tabular}{ccccccc}
\hline
$\hat{\Lambda}$ & $n$ & $E_{extractable,n}/M_0$ & $E_{extracted,n}/M_0$ & $M_{irr,n}/M_0$ & $\xi_n$ & $\Xi_n$ \\
\hline
\multirow{9}{*}{0} 
& 0 & 0.292893 & 0.000000 & 0.707107 & 0.000000 & 0.000000 \\
& 1 & 0.275611 & 0.001450 & 0.722940 & 0.144958 & 0.083875 \\
& 2 & 0.268419 & 0.002899 & 0.728682 & 0.144960 & 0.118460 \\
& 3 & 0.262915 & 0.004349 & 0.732736 & 0.144962 & 0.145066 \\
& 4 & 0.258294 & 0.005799 & 0.735907 & 0.144965 & 0.167593 \\
& 5 & 0.254245 & 0.007248 & 0.738506 & 0.144967 & 0.187549 \\
& 6 & 0.250608 & 0.008698 & 0.740694 & 0.144969 & 0.205702 \\
& 7 & 0.247286 & 0.010148 & 0.742566 & 0.144971 & 0.222510 \\
& 8 & 0.244218 & 0.011598 & 0.744184 & 0.144974 & 0.238271 \\
\hline
\multirow{7}{*}{0.03}
& 0 & 0.288388 & 0.000000 & 0.722388 & 0.000000 & 0.000000 \\
& 1 & 0.272711 & 0.001470 & 0.737534 & 0.147048 & 0.093798 \\
& 2 & 0.266176 & 0.002941 & 0.742960 & 0.147048 & 0.132402 \\
& 3 & 0.261172 & 0.004411 & 0.746757 & 0.147048 & 0.162088 \\
& 4 & 0.256972 & 0.005882 & 0.749698 & 0.147048 & 0.187225 \\
& 5 & 0.253293 & 0.007352 & 0.752084 & 0.147048 & 0.209498 \\
& 6 & 0.249990 & 0.008823 & 0.754068 & 0.147048 & 0.229774 \\
\hline
\multirow{6}{*}{0.06}
& 0 & 0.282961 & 0.000000 & 0.740547 & 0.000000 & 0.000000 \\
& 1 & 0.269123 & 0.001490 & 0.754740 & 0.148989 & 0.107670 \\
& 2 & 0.263350 & 0.002980 & 0.759720 & 0.148986 & 0.151943 \\
& 3 & 0.258933 & 0.004470 & 0.763145 & 0.148984 & 0.186018 \\
& 4 & 0.255232 & 0.005959 & 0.765750 & 0.148982 & 0.214918 \\
& 5 & 0.251999 & 0.007449 & 0.767816 & 0.148980 & 0.240588 \\
\hline
\multirow{2}{*}{0.099555}
& 0 & 0.273542 & 0.000000 & 0.771447 & 0.000000 & 0.000000 \\
& 1 & 0.262820 & 0.001513 & 0.783468 & 0.151302 & 0.141110 \\
\hline
\end{tabular}
\end{table}
From Table \ref{tab:2}, it can be observed that the EROI $\xi$ for each Penrose process is nearly identical, and a larger cosmological parameter results in a larger value of $\xi$, once again indicating that a larger cosmological parameter enhances the single-extraction energy capability. Furthermore, it can be seen that for the same number of iterations, a larger cosmological parameter leads to higher values of $\Xi$ and $E_{extracted,n}/M_0$, further demonstrating that Kerr-dS spacetime possesses a higher single-extraction energy capability compared to Kerr spacetime. However, due to the influence of the iteration stopping condition, at a lower decay radius, a larger cosmological parameter causes the process to stop sooner. This results in a lower EUE and less total extracted energy after iteration termination. For example, for a decay radius $\hat{r}_x=1.2$, a Kerr black hole with initial $\hat{\Lambda}=0$ can iterate 8 times, while with $\hat{\Lambda}=0.03$ and $0.06$, it can only iterate 6 and 5 times, respectively. For $\hat{\Lambda}=0.099555$, it can only iterate once. In contrast, at a higher decay radius, e.g., $\hat{r}_x=1.8$, the initial values $\hat{\Lambda}=0, 0.03, 0.06, 0.099555$ can iterate 30, 34, 36, and 36 times, respectively. Coupled with the fact that a larger cosmological parameter corresponds to a greater single-extraction energy capability, it is easy to understand why, after iteration termination and at higher decay radii, a larger cosmological parameter  yields a  higher EUE and more extracted energy, as shown in the middle and right panels of Fig. \ref{fig:7}. However, it is noteworthy that the EUE never exceeds $50\%$. Furthermore, at a higher decay radius, the EUE is even lower for the same cosmological parameter. This indicates that in the repetitive Penrose process, the change in extractable energy is primarily converted into irreducible mass, this law still holds true for Kerr-dS black holes.

\section{Conclusion}

In this paper, we have studied the repetitive Penrose process in extremal Kerr-dS spacetime. Firstly, we provided a brief review of Kerr-dS spacetime. Under the optimal conditions for maximum energy extraction, we derived an analytical solution for the fundamental equations of the Penrose process on the equatorial plane within a general metric framework. Then, we introduced the iterative procedure and the conditions that must be satisfied. In particular, by plotting the variation of $\hat{a}_{\min}$ for the three particles with respect to $\hat{r}_x$ for different values of $\Lambda M^2$, we concluded that the lower spin limit  stopping the iteration is determined by particle 0. Finally, we presented the results of this process.

The results show that the repetitive Penrose process cannot extract all the extractable energy from Kerr-dS black holes. Only a small portion of the reduction in extractable energy is converted into extracted energy, while the majority is transformed into irreducible mass. As the iteration is ended, a significant amount of extractable energy remains, and it may be extracted through other means, which aligns with previous conclusions. Furthermore, the Kerr-dS spacetime exhibits a higher EROI and a greater single-extraction energy capability compared to the Kerr spacetime. The larger the cosmological parameter, the stronger the EROI and single-extraction energy capability. In cases with a lower decay radius, after the completion of the repetitive Penrose process, the Kerr black hole demonstrates a higher EUE and yields more extracted energy. However, in cases with a higher decay radius, the situation is reversed, the Kerr-dS black hole shows a higher EUE and yields more extracted energy. This is due to the existence of an iteration stopping condition.

\noindent {\bf Acknowledgments}

\noindent
This work is supported by the National Natural Science Foundation of China (Grants Nos. 12375043,
12575069 ).

\end{document}